\documentclass[aps,floatfix,nofootinbib,amsmath,amssymb,showpacs]{revtex4}

\usepackage[dvips]{graphicx}
\usepackage{mathrsfs}
\usepackage{slashed}

\newcommand{\be}{\begin{equation}}
\newcommand{\ee}{\end{equation}}

\begin{document} 

\preprint{UCI-TR-2010-01}

\title{Maverick dark matter at colliders}

\author{Maria Beltr\'an}
\affiliation{Department of Astronomy and 
Astrophysics and Enrico Fermi Institute, University of Chicago, 5640 S.\ Ellis
Ave., Chicago, Illinois 60637}

\author{Dan Hooper}
\affiliation{Theoretical Astrophysics, Fermi National Accelerator Laboratory,
Box 500, Batavia, Illinois 60637}
\affiliation{Department of Astronomy and Astrophysics and Kavli Institute for
Cosmological Physics, University of Chicago, 5640 S.\ Ellis
Ave., Chicago, Illinois 60637}

\author{Edward W.\ Kolb}
\affiliation{Department of Astronomy and Astrophysics, 
Enrico Fermi Institute, and Kavli Institute for Cosmological Physics, 
University of Chicago, 5640 S.\ Ellis Ave., Chicago, Illinois 60637}

\author{Zosia A.\ C.\ Krusberg}
\affiliation{Department of Physics and Enrico Fermi Institute, 
University of Chicago, 5640 S.\ Ellis Ave., Chicago, Illinois 60637}

\author{Tim M.\ P.\ Tait}
\affiliation{Department of Physics and Astronomy, University of California, 
Irvine, California 92697}


\begin{abstract}

Assuming that dark matter is a weakly interacting massive particle (WIMP)
species $X$ produced in the early Universe as a cold thermal relic, we study the
collider signal of $pp$ or $p\bar{p}\rightarrow\bar{X}X + \textrm{jets}$ and its
distinguishability from standard-model background processes associated with jets
and missing energy.  We assume that the WIMP is the sole particle related to
dark matter within reach of the LHC---a ``maverick'' particle---and that it
couples to quarks through a higher dimensional contact interaction.   We
simulate the WIMP final-state signal $X \bar{X} + \textrm{jets}$ and dominant
standard-model (SM) background processes and find that the dark-matter
production process results in higher energies for the colored final state
partons than do the standard-model background processes.  As a consequence, the
detectable signature of maverick dark matter is an excess over standard-model
expectations of events consisting of large missing transverse energy, together
with large leading jet transverse momentum and scalar sum of the transverse
momenta of the jets. Existing Tevatron data and forthcoming LHC data can
constrain (or discover!) maverick dark matter.

\end{abstract}
\pacs{13.85.Rm,14.80.-j,98.80.Cq,95.35.+d}
\maketitle

\section{Introduction}

Compelling observational evidence for the existence of dark matter has been
found on a wide range of astronomical scales.  However, only a few things are
presently known about the physical properties of dark matter: its relic density
has been determined very precisely by the WMAP experiment to be
$\Omega_Xh^2=0.1109 \pm 0.0056$ \cite{Komatsu:2010fb}, and observations suggest
that it is non-baryonic, cold, dissipationless, and stable on time scales on the
order of the age of the universe.  Many dark-matter candidates have been
proposed.  Among these, weakly interacting massive particles (WIMPs) are
particularly compelling: they are predicted by many theories of beyond the
standard-model (BSM) physics and can naturally produce the observed relic
dark-matter density through thermal processes.

Many previous studies have sought to constrain the properties of various WIMP
candidates using collider data.  Most of these have studied WIMP candidates
within specific theoretical frameworks such as supersymmetry \cite{neutralino}
or models with warped \cite{warped} or universal \cite{universal}
extra dimensions.  The constraints have normally been imposed by various
collider bounds, such as the bounds on the Higgs mass, the invisible $Z$ width,
the masses of new charged and colored particles, the masses of new charged and
neutral gauge bosons, flavor-changing neutral currents, the branching ratios of
$b \rightarrow s\gamma$ and $B_s \rightarrow \mu^+ \mu^-$, the anomalous
magnetic moment of the muon, and other electroweak precision
measurements \cite{bertone05}.  Stringent constraints on various particle models
have been obtained in such analyses (see \textit{e.g.,} Ref.\ \cite{baer03} for
an analysis of mSUGRA parameters and Ref.\ \cite{cdf06} for an analysis of large
extra dimensions parameters).

The theoretical frameworks to which these WIMP candidates belong are often both
theoretically well motivated and compelling.  However, given that all of these
theories still lack experimental support, we cannot exclude the possibility that
dark matter belongs to some other, yet unidentified, theory.  Additionally,
given that the first observations of dark matter may come from direct- or
indirect-detection experiments, which may only provide information about the
general properties of the dark-matter particle without offering a way to
distinguish between underlying theories, it is important to remain unbiased
about the nature of dark matter.  For this reason, model-independent studies of
dark-matter phenomenology using effective field theory can be particularly
important.  

In fact, examining the primary  experimental analyses of our favorite theories
reveals a common feature of such constraints: they are relatively insensitive to
the nature of the dark-matter  particles \textit{per se,} and instead probe the
properties of the new, exotic colored and/or charged states that accompany the
dark-matter particle.  Such states are more amenable to experimental probes
because they have large standard-model gauge interactions and are thus more
easily produced.  However, if our primary motivation is to learn about the
nature of dark matter, they are something of a distraction.  Further, when such
charged states are somewhat heavier than the dark matter itself, the only
signals accessible to colliders (at least in early data sets) may be the
production of new, heavy SM-charged particles.

In a recent paper \cite{beltran09}, an effective field theory approach was used
to evaluate the constraints on generic WIMP candidates from direct-detection
experiments and the prospects for a signal in future indirect-detection
experiments.  Several WIMP candidates with specified spin and interaction forms
with standard-model fermions were found to be excluded on the basis of
direct-detection experiments.  An interesting possibility---a fermionic WIMP
with an axial-vector coupling to standard-model fermions---was not ruled out by
direct-detection experiments, and products of its annihilation in the Sun are
expected to be eventually detectable by a neutrino detector such as IceCube.  
Motivated by this result, we study dark matter at accelerators by investigating
the production and detection prospects of this WIMP candidate at the Tevatron
and at the LHC.  

Our effective field theory description of dark matter is largely model
independent, but differs in important ways from previous model-independent
analyses \cite{Birkedal:2004xn,Feng:2005gj}.  These previous works take the
annihilation cross section as inferred from the relic density, time-reverse it
into a collider production process (taking into account subtleties related to
spin and flux factors appropriately), and use factorization theorems to produce
SM radiation.  Our effective field theory description pays the price of being
more model-dependent, but has the advantage that it can describe relativistic
production, which may differ substantially from the nonrelativistic cross
section relevant to describe freeze out.  This difference is particularly
important if the nonrelativistic cross section appropriate for the freeze-out
calculation is $p$-wave suppressed, as in the case of the axial-vector
interactions we will study. In fact, this difference will contribute to our
somewhat more optimistic conclusions concerning the potential for observing
WIMPs at the LHC.

\section{Effective field theory\label{EFT}}

We assume that our WIMP candidate is the only new particle species within reach
of the LHC (a ``maverick'' particle); this allows us to describe its interaction
with standard-model quarks accurately in terms of an effective field theory
whose degrees of freedom consist of the standard-model (SM) particles plus the
WIMP itself.  For the purposes of the discussion, we also specialize to the case
of a Dirac fermion. The effective field theory consists of the SM Lagrangian
plus kinetic terms for the dark matter $X$ and a set of effective four-Fermion
interactions between $X$ and the quarks $q = u, d, s, c, b, t$, 
\be 
\mathcal{L} =  \mathcal{L}_{SM} + i \bar{X}
\gamma^\mu \partial_\mu X - M_X \bar{X} X +  \sum_q
\sum_{i,j}{\frac{G_{qij}}{\sqrt{2}} ~  \left[ \bar X \Gamma_i^X X \right] 
\left[\bar q \Gamma^j_q q \right] }, 
\label{EFT1} 
\ee 
where the sums $i,j$ are over scalar, pseudoscalar, vector, axial vector, and
tensor interactions (in Lorentz-invariant combinations) described by the
operators $\{\Gamma\}$.  We will assume that the interaction is dominated by
only one of the above forms.\footnote{As in Ref.\ \cite{beltran09}, we do not
consider WIMP  couplings to Higgs or gauge bosons.  More general actions
including such terms can be found in Ref.\ \cite{Shepherd:2009sa}.}  The mass
dimension of $G_{qij}$ is minus two.  Although it would be straightforward to
account for WIMPs coupling to leptons,  for simplicity, we do not  take this
possibility into account in our analysis, and we further specialize to the case
in which the WIMP couplings are independent of the quark flavor, $G_{qij} =
G_{ij}$. With these assumptions, and the assumption that the dark-matter density
is determined by the calculation of the thermal relic abundance of the $X$s, the
various $G_{i}$'s are determined to high precision.  Even if the WIMP proves not
to be a thermal relic, the effective field theory (with the $G$'s uncorrelated
from $M_X$) may still provide a useful language to describe WIMP coupling to SM
particles.

\begin{figure}
\centering
\includegraphics[width=5in]{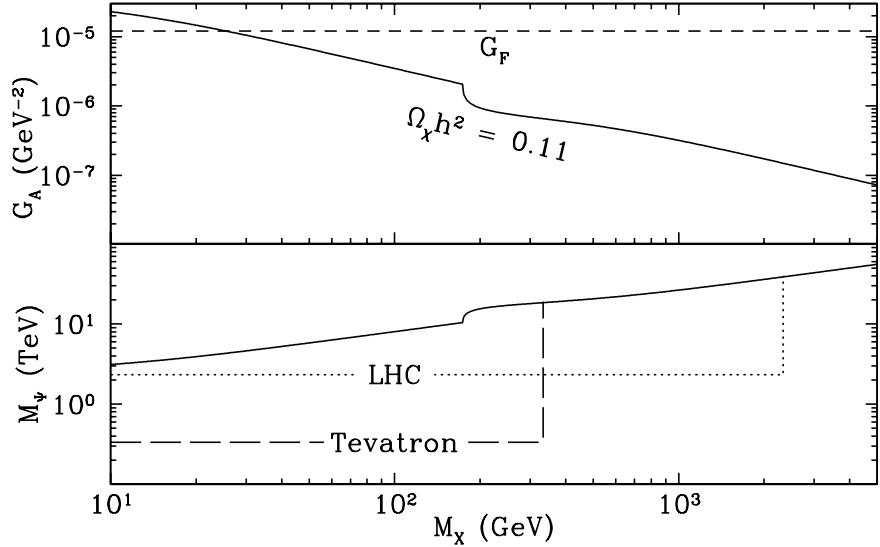}

\caption{The upper panel indicates the value of $G_A$ as a function of the WIMP
mass $M_X$ necessary to result in a relic abundance of $\Omega_X h^2 = 0.11$ if
the interaction Lagrangian is given by Eq.\ (\ref{interactionlagrangian}). 
Shown for comparison is the value of the Fermi coupling constant, $G_F$. For the
effective field theory of Sect.\ \ref{EFT} to be valid, the mass of $\Psi_\mu$
in Eq.\ (\ref{uvcomplete}) must be below the solid curve and above the dashed
(Tevatron) and dotted (LHC) horizontal lines in the bottom panel.  The vertical
dashed and dotted lines indicate the rough kinematic reach for dark-matter
production for the indicated hadron collider.  The kinks in the curves around 
$M_X = 170$ GeV correspond to the opening of the top quark production channel, 
allowing for weaker couplings (top panel) and consequently requiring larger $M_{\psi}'s$
(bottom panel).}

\label{mm}
\end{figure}

In the case where the WIMP interaction is spin independent, the allowed range
for the WIMP mass and coupling constant is very tightly constrained from
direct-detection search experiments \cite{beltran09}. However if the
$Xq\rightarrow Xq$ scattering is \textit{spin dependent}, then the corresponding
direct-detection limits are too weak to exclude any of the acceptable WIMP mass 
and coupling constant combinations. Therefore, we will consider interactions
that yield a spin-dependent scattering cross section.  In particular, we 
simplify the consideration to the axial-vector case and consider a Lagrangian
\be
\mathcal{L}_{int} = \frac{G_A}{\sqrt{2}} ~\sum_q
\left[ \bar X \gamma^{\mu} \gamma^5 X \right]
\left[ \bar q \gamma_{\mu} \gamma_5 q \right] .
\label{interactionlagrangian}
\ee
The coupling constant $G_A$ is obtained for a WIMP of a given mass from the
requirement that it leads to the correct relic density found by WMAP within the
standard thermal freezeout framework (see, \textit{e.g.,} Ref.\ 
\cite{kolbturner}).  The resulting $G_A$ as a function of $M_X$ is shown in
the upper panel of Fig.\ \ref{mm}.  Our primary
assumption that the WIMP is the only new particle within LHC reach also
ensures that there are no resonances or coannihilations to complicate the
standard analysis of the relic density.

While we work with an effective low-energy field theory, it is instructive to
discuss possible ultraviolet (UV) completions of our theory.  One possible
completion would involve a massive intermediate vector boson $\Psi_\mu$ with
mass much larger than any other mass or energy scale associated with the
low-energy effective field theory.  Then $G_A$ of the effective field theory is
related to parameters of the UV complete theory by
\be 
\frac{G_A}{\sqrt{2}} = \frac{g_q g_X}{M_\Psi^2}, 
\label{uvcomplete} 
\ee 
where $g_q$ ($g_X$) is the coupling constant for the $\Psi$--$\bar{q}$--$q$ 
($\Psi$--$\bar{X}$--$X$) interaction. With the requirement that the
dimensionless coupling constants $g_q$ and $g_X$ are smaller than $4\pi$, for a
given value of $G_A(M_X)$ (set by the relic density) there is a maximum value of
$M_\Psi$ such that the UV completion admits a perturbative description.  This is
shown by the solid curve in the bottom panel of Fig.\ \ref{mm}.   For the
effective theory to make sense, $\Psi_\mu$ must be massive enough so as to not
be directly accessible at the energies of interest.  At a hadron collider of
center of mass energy $E$, this requirement ultimately depends on the details of
the relevant parton distribution functions (PDFs) and the collected luminosity. 
A very rough estimate is furnished by $M_\Psi \gtrsim E/3$.  The  resulting
lower limits for the Tevatron and the LHC are shown by the horizontal lines in
the bottom panel of Fig.\ \ref{mm}. Also shown as the vertical lines in the
bottom panel of Fig.\ \ref{mm} is the approximate kinematic mass reach for $X
\bar{X}$ production at the Tevatron and LHC.  We will discuss detailed estimates
for the discovery potential of $X$ as a function of its mass in the sections
below.

\section{Collider Signal and Background \label{sec:collider}}

\subsection{Processes}

Our effective theory for WIMP -- SM interactions leads to the production of
WIMPs at colliders through the process
\be
pp \ (p\bar{p} ) \rightarrow X \bar{X} .
\ee
However, this process is worthless as a discovery mode at a hadron collider
because it contains no visible trace that a hard scattering took place at all. 
Consequently, we turn to the process in which WIMPs are produced together with a
hard  parton,\footnote{Recently, Ref.\ \cite{Cao:2009uw}  studied WIMPs produced
at the LHC together with a hard photon.}
\be
pp \ (p\bar{p} ) \rightarrow X \bar X + \textrm{jets} .
\label{signal}
\ee
While this process is formally higher order in perturbation theory, the hard
jet(s) of  hadrons provides a trigger that a hard scattering actually took
place, with the  WIMPs ``seen" as missing momentum against which the jet
recoils.

The dominant SM physics backgrounds consist of electroweak
processes, such as $Z\ + $ jets, where the $Z$ decays into a pair of neutrinos, 
\be
pp \ ( p\bar{p}) \rightarrow \nu \bar \nu + \textrm{jets},
\label{background1}
\ee
as well as $W^\pm\ +$ jets where the $W$ decays into a neutrino and a charged
lepton, 
\be
pp \ (p\bar{p}) \rightarrow 
l^- \bar \nu + \textrm{jets}~~\textrm{and}~~p p \ (p\bar{p})
\rightarrow l^+ \nu + \textrm{jets} ,
\label{background2}
\ee
and the charged lepton either falls outside of the acceptance range of the
detector or is lost inside a jet.  At the LHC, we also consider the background
from $t \bar{t}$ production:
\be
p p \rightarrow t \bar{t} \rightarrow W^+ b~ W^- \bar{b} ,
\ee
whose decays again produce $W$ bosons. There are additional ``QCD" backgrounds
that arise from purely strong-interaction processes in which mismeasurement
leads to fake missing transverse momentum.  This background depends intricately
on the details of the detector, and is beyond the scope of our ability to
model properly. That said, we will apply stiff missing momentum cuts and require
the leading jet to be acollinear with the missing transverse momentum. Both cuts
should help minimize the sensitivity to detector details on our search proposal.

\begin{figure}
\centering
\includegraphics[width=2in]{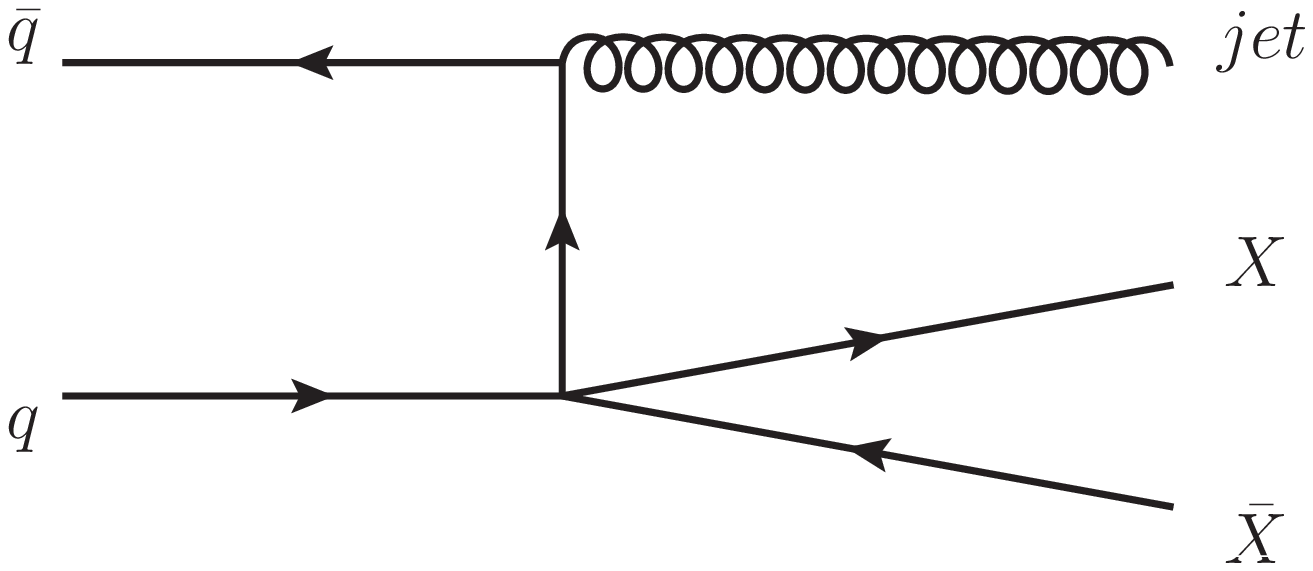}
\includegraphics[width=2in]{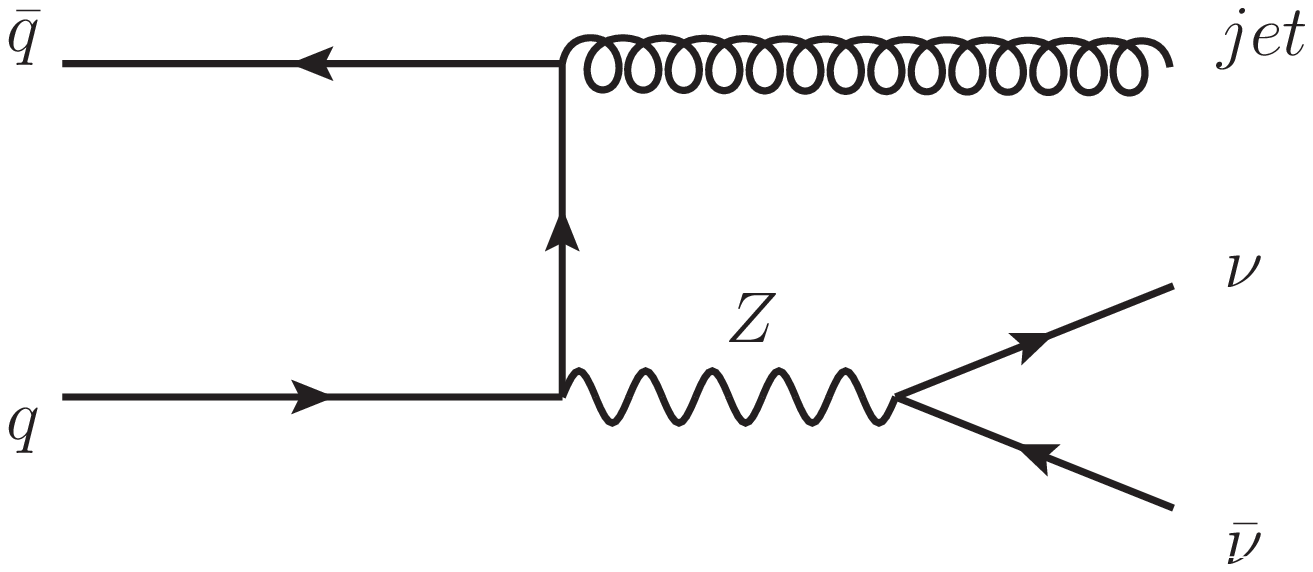}
\includegraphics[width=2in]{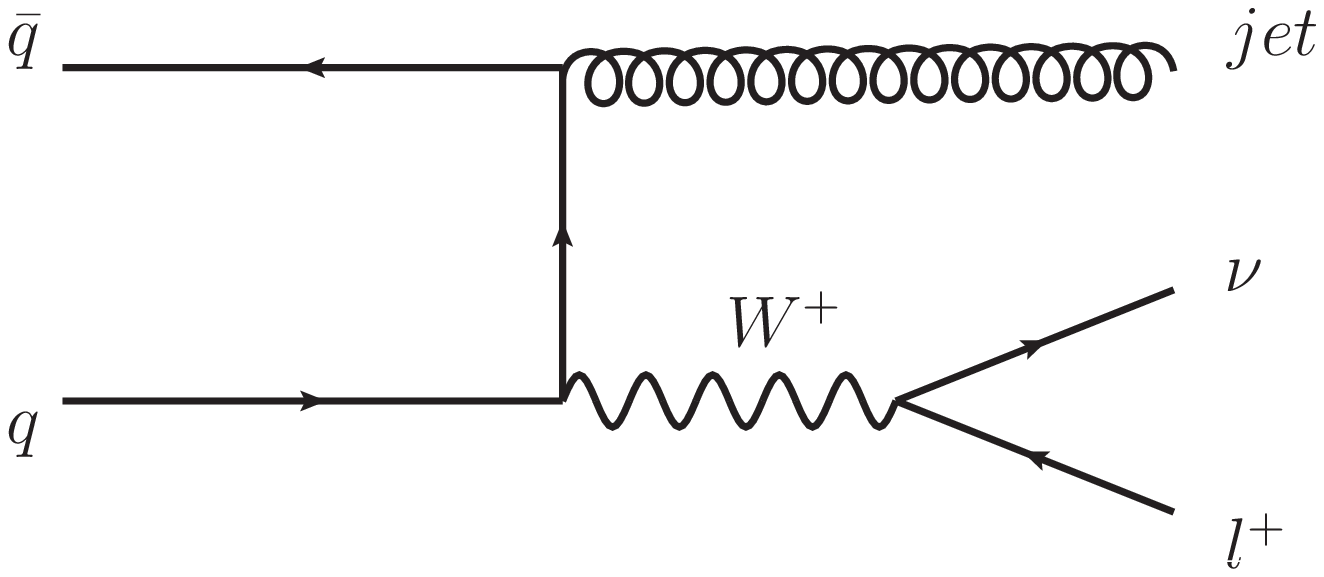}

\caption{Representative Feynman diagrams (at the parton level) for the processes
$pp\rightarrow X\bar{X}+\textrm{ jet}$ (left), $pp\rightarrow \nu\bar{\nu}
+\textrm{ jet}$ (center), and $pp\rightarrow l^+\nu+\textrm{ jet}$ (right). } 

\label{Feynman}
\end{figure}

Representative parton-level Feynman diagrams for signal and electroweak
background processes are shown in Fig.\ \ref{Feynman}.  We simulate the signal
and background events using the MadEvent package \cite{madgraph}, in which we
have implemented a Dirac fermion WIMP which interacts with the SM through the
maverick interaction of Eq.\ (\ref{interactionlagrangian}). After MadEvent
generates the hard scattering process,  Pythia \cite{pythia} is called to
simulate parton showering and hadronization, and PGS (with the generic Tevatron
and LHC detector models) provides an estimate of the detector effects
\cite{pgs}. 

While the dominant corrections to the kinematics from higher orders of
perturbation theory are captured by the parton shower, higher order
contributions also correct the over-all rates of the processes.  We improve our
estimates by applying a  flat $K$-factor to the SM background rates,
\be
K\textrm{-factor} = \frac{\sigma_{\textrm{NLO}}}{\sigma_{\textrm{LO}}}.
\ee
A flat $K$-factor is known to work reasonably well for both $W$ + jets and $Z$ +
jets, and their values are similar for both processes \cite{campbell07},
\begin{align}
\textrm{Tevatron:~}&K_{W,Z} = 1.20 \\
\textrm{LHC:~}&K_{W,Z} = 1.32. 
\end{align}
While we expect that a similar $K$-factor applies to the WIMP signal process, in
the absence of an explicit computation and to be conservative, we use leading
order rates for the signal.  The $K$-factor for $t \bar{t}$ at the LHC is taken
to be \cite{Beenakker:1990maa}
\be
K_{t \bar{t}} = 1.54 ~.
\ee

\begin{table*}[!ht]
\hspace{0.0cm}
\begin{ruledtabular}
\begin{tabular}{l c c c c}
Collider & Process & $M_\chi$ (GeV) & Cross section before cuts (pb)
& Cross section after cuts (pb) \\ 
\hline\hline
Tevatron & $p \bar p \rightarrow X \bar X J$ & 5 & $4.18 \times 10^1$ 
& $2.43 \times 10^{-2}$ \\
 & & 10 & $1.74 \times 10^1 $ & $1.05 \times 10^{-2}$ \\
 & & 25 & $4.16 \times 10^0 $ & $2.25 \times 10^{-3}$ \\
 & & 50 & $9.62 \times 10^{-1}$ & $9.62 \times 10^{-4}$ \\
 & $p \bar p \rightarrow \nu \bar \nu J$ & --- & $1.64 \times 10^2$ 
 & $1.15 \times 10^{-2}$ \\
 & $p \bar p \rightarrow l^- \bar \nu J$ & --- & $4.19 \times 10^2$ & 
 $ 4.20\times10^{-3}$ \\
 & $p \bar p \rightarrow l^+ \nu J$ & --- & $4.20 \times 10^2$ & 
 $8.43 \times 10^{-3}$ \footnotemark \\
\hline 
LHC & $pp \rightarrow X \bar XJ$ & 10 & $9.42 \times 10^2$ & 
$1.84 \times 10^1$ \\
 & & 50 & $6.47 \times 10^1$ & $1.27 \times 10^0$ \\
 & & 100 & $1.51 \times 10^1$ & $3.30 \times 10^{-1}$ \\
 & & 500 & $1.18 \times 10^{-1}$ & $3.82 \times 10^{-3}$ \\
 & $pp \rightarrow \nu \bar \nu J$ & --- & $2.52 \times 10^3$ 
 & $4.16 \times 10^{-1}$ \\
 & $pp \rightarrow l^- \bar \nu J$ & --- & $4.00 \times 10^3$ 
 & $7.05 \times 10^{-2}$ \\
 & $pp \rightarrow l^+ \nu J$ & --- & $5.26 \times 10^3$ 
 & $1.68 \times 10^{-1}$ \\
 & $pp \rightarrow t \bar{t}$ & --- & $7.98 \times 10^2$
 & $5.10 \times 10^{-2}$ \\
\end{tabular}
\end{ruledtabular}

\footnotetext{Because $W$ interactions are left-handed (and the light fermions are 
 effectively massless), there are different spin correlations in the $p_T$ distributions for $l^+$ and $l^-$. 
 Charged leptons produced by $W^+$'s are anti-particles, with right-handed helicity, while charged leptons 
 produced by $W^-$'s are particles, with left-handed helicity.  The difference in the $p_T$ distributions 
 results in different efficiencies to pass our cuts that veto high-$p_T$ leptons.  (This also applies to the 
 cross sections for the analogous processes at the LHC.)}

\caption{Cross sections for the production of $X\bar{X}$+jets for the indicated
WIMP masses and the background processes  specified in Eqs.\
(\ref{background1})--(\ref{background2}) at the Tevatron (center-of-mass energy
$\sqrt{s}=1.96$ TeV) and the LHC (center-of-mass energy $\sqrt{s}=14$ TeV),
before and after cuts.}

\label{crosssections1}
\end{table*}

Inclusive cross sections for signal and background processes at the Tevatron  (a
$p \bar{p}$ collider with $\sqrt{s} = 1.96$ TeV) and the LHC (a $p p$ collider
with $\sqrt{s} = 14$ TeV) are given in Table \ref{crosssections1}.  Minimal cuts
of $p_T \geq 20$~GeV and $| \eta | \leq $ 3.6 (Tevatron) $|\eta| \leq $ 2.5
(LHC) are imposed on the hard parton, in order to render the rates IR-safe.  
The quantity $\eta$ is the pseudo-rapidity of the jet. Such loose cuts are
marginally realistic at the Tevatron and completely unrealistic at the LHC.  We
will discuss our actual (realistic) analysis cuts below.

\begin{figure}
\centering
\includegraphics[width=5in]{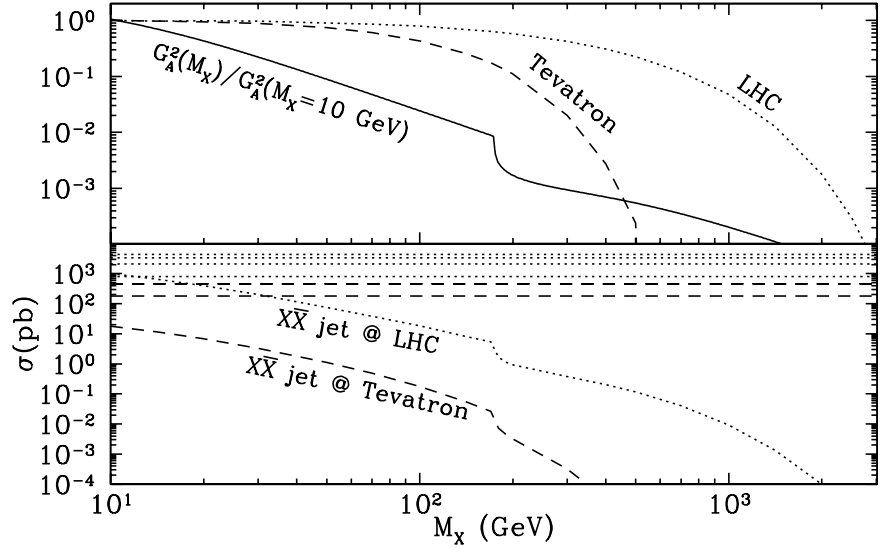}

\caption{Upper panel: Suppression factors as a function of the WIMP mass.  The
broken curves track the kinematic and parton flux suppression for the indicated
hadron colliders, normalized to unity at a mass of 10 GeV.  The solid curve
represents the suppression due to the decrease in $G_A$ as $M_X$ increases
(again normalized to unity at a mass of 10 GeV).  The total suppression as a
function of mass is the product of the two factors.  Lower panel: $X \bar{X}$ +
jet signal cross section as a function of the $X$ mass and the cross sections
for the background processes at the LHC (dottel lines) and the Tevatron (dashed
lines).The background processes for the indicated colliders are, from top to
bottom, $l^+$ jet, $l^-$ jet, $\nu\bar{\nu}$ jet, and $t\bar{t}$. (The cross
sections for $l^+$ jet and $l^-$ jet at the Tevatron are nearly
indistinguishable.)   The kinks in the curves around $M_X = 170$ GeV correspond to the
opening up of the top quark production channel, allowing for weaker couplings (top panel) 
and correspondingly lower cross sections (bottom panel).} 

\label{sig}
\end{figure}

As the $X$ mass increases, the signal from $X \bar{X}$ + jet decreases.  This is
in part because we adjust $G_A$ together with $M_X$ to hold the thermal relic
density fixed. The relic density is inversely proportional to the cross section
for $X \bar{X} \rightarrow q \bar{q}$, which scales roughly as $\sigma \propto
G_A^2M_X^2$, so keeping the relic density fixed implies $G_A(M_X)\propto
M_X^{-1}$.  In addition, higher energy partons are required to produce heavier
$X$ particles, and the parton distribution functions fall steadily  with the
energy of the parton. The separation of the two effects is illustrated in Fig.\
\ref{sig}, along with the cross sections as a function of $M_X$.  

\subsection{Analysis and Cuts}

\begin{figure}
\centering
\includegraphics[width=3in]{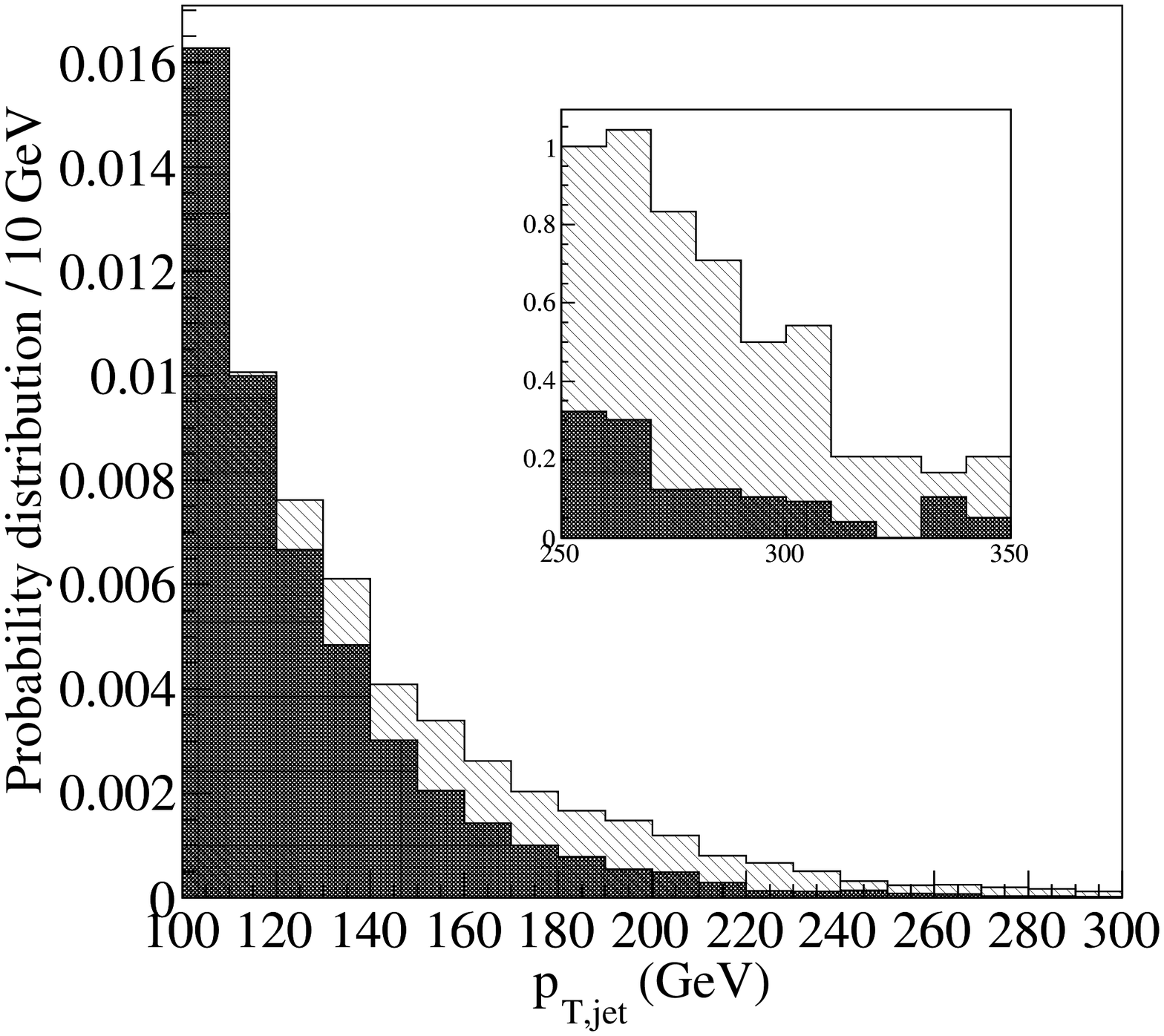}
\includegraphics[width=3in]{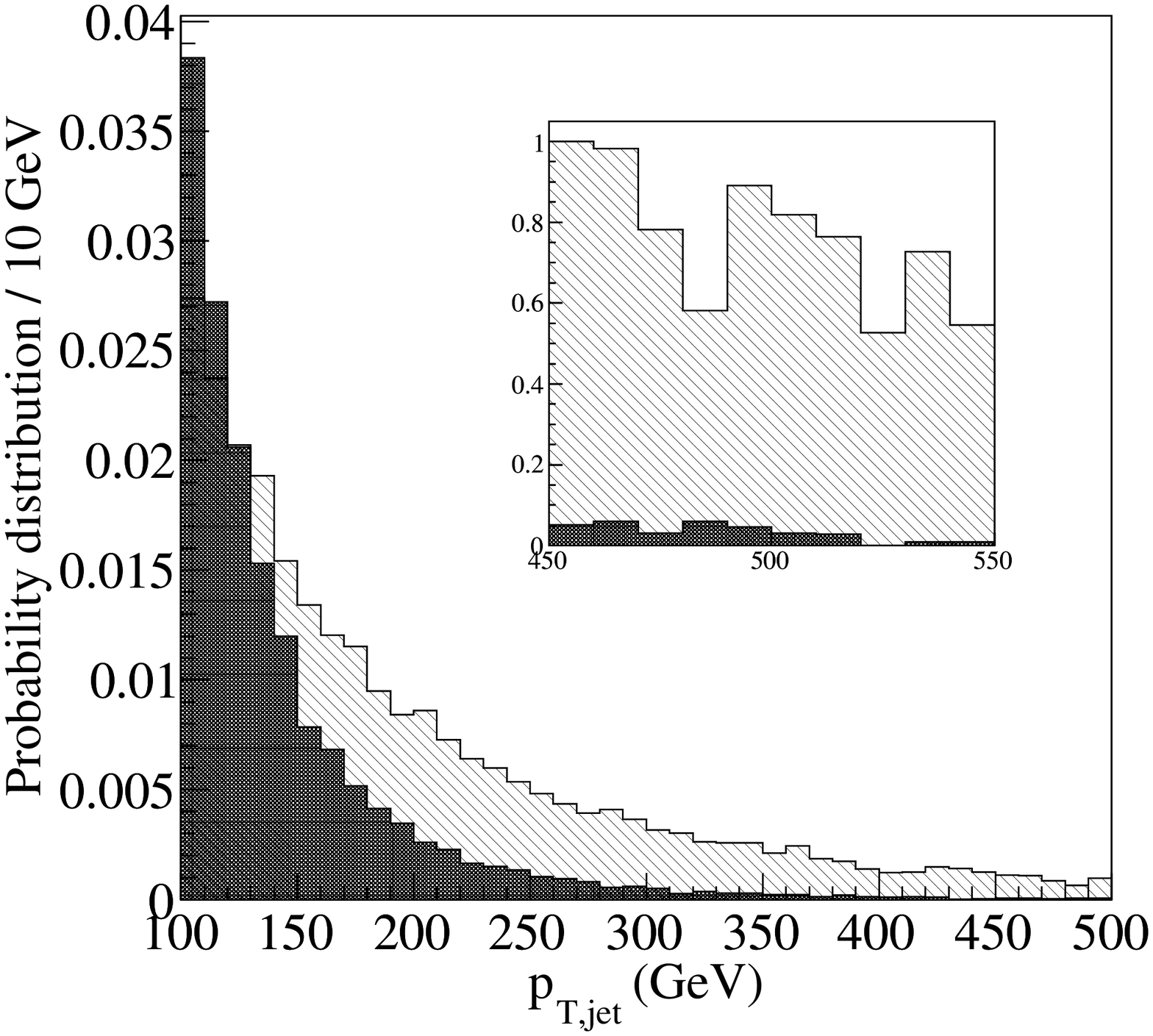}

\caption{The normalized distributions of the transverse momentum of the leading
\textit{parton} for the standard-model background (dark region) and the maverick
dark-matter signal (light region) at the Tevtron (left) and the LHC (right).  
The insert in each the figure shows the distribution of  events for which the
leading parton has transverse momentum satisfying  $p_{T} \geq
p_{T,\textrm{min}}$.  The signal has been generated with $M_X=5$ GeV at the
Tevatron and $M_X=50$ GeV at the LHC.} 

\label{parton}
\end{figure}

Our missing energy signature typically contains only jets as visible particles
in the detector.  In order to trigger on the events, we require that the leading
jet is central and has a minimum transverse momentum.  These cuts are 
$p_{T,\textrm{jet}} > 250$ GeV, $| \eta | \leq$ 3.6 at the Tevatron or
$p_{T,\textrm{jet}} > 450$ GeV, $| \eta | \leq$ 2.5 at the LHC. The $p_T$ cuts
are somewhat greater than those sufficient for an inclusive jet trigger at CDF
and D$\slashed{0}$ at the Tevatron, or ATLAS and CMS at the LHC.  In order to
select events in which WIMPs may have been produced, we require a missing energy
cut of $\slashed{E}_T > 30$ GeV (both colliders).  To minimize the contribution
of fake missing energy from jet mismeasurements, we require that the
$\slashed{E}_T$ and leading jet point in different azimuthal directions,  $|
\phi_{\textrm{jet}} - \phi_{\slashed{E}_T}| > 0.5$~rad \cite{Ball:2007zza}.

In order to reduce backgrounds containing missing energy from processes where
$W$ bosons are produced, we veto events containing one or more energetic
isolated charged leptons (either $e^\pm$, $\mu^\pm$, or hadronically
reconstructed $\tau^\pm$).  We remove such events for which a charged lepton has
transverse momenta $p_{T,l^{\pm}} \geq 10$ GeV and is further separated from all
jets by  $\Delta \mathcal{R} = \sqrt{(\Delta \eta)^2+(\Delta\phi)^2} > 0.4$.
Softer leptons may be inefficiently reconstructed by the detector, and those
closer to jets may be ``lost" inside them or result from heavy flavor decays
inside the jet hadrons.  Our implementation of the charged lepton cuts is very
similar to those used by a recent CDF monojet search \cite{Abulencia:2006kk},
and we are able to reproduce the post-cut event number quoted by CDF. In
addition, to help with the $t \bar{t}$ background, we reject events in which one
or more of the jets is tagged as containing a bottom quark.

Maverick dark matter can interact with SM particles only via non-renormalizable
interaction terms.  Such terms produce cross sections which fall less slowly
compared to those mediated entirely by SM interactions.  Thus, we expect the
energy of the associated jet in $X \bar{X}+$ jet will typically be more than
the energy associated with $W+$ jet or $Z+$ jet, providing a handle that 
potentially can be used to distinguish the signal from the background.   In
Fig.\ \ref{parton}, we plot (at the parton level) the $p_T$ of the parton
produced in the hard scattering in events from $X \bar{X}+$ jet and the SM
backgrounds.  The distributions are normalized to unity to  emphasize the
difference in the shapes with respect to the parton $p_T$. 

Corrections from higher orders in perturbative QCD may modify the energy 
distribution of the hard parton. While the parton shower correctly captures
higher order corrections to the kinematics resulting from soft or collinear
radiation, it fails when the additional parton is radiated at wide angles with
respect to its parent.  Improved simulation of the kinematic distributions is
possible by matching Monte Carlo samples with exact leading order matrix
elements for additional radiation  in the hard scattering process to those for
lower order processes.  These ``jet matched samples" have the virtue of using
the parton shower to resum soft and collinear logarithms while incorporating the
exact matrix elements for perturbative radiation \cite{Catani:2001cc}.  In
missing energy observables, they tend to enhance tails in the missing energy
distributions of both signal and background processes \cite{Alwall:2008ve}.
While a detailed simulation of such effects is beyond the scope of this work, we
expect it will shift both the signal and background $\slashed{E}_T$
distributions by a similar factor, and should ultimately be included when
analyzing collider data.

\begin{figure}
\centering
\includegraphics[width=3.5in,angle=0]{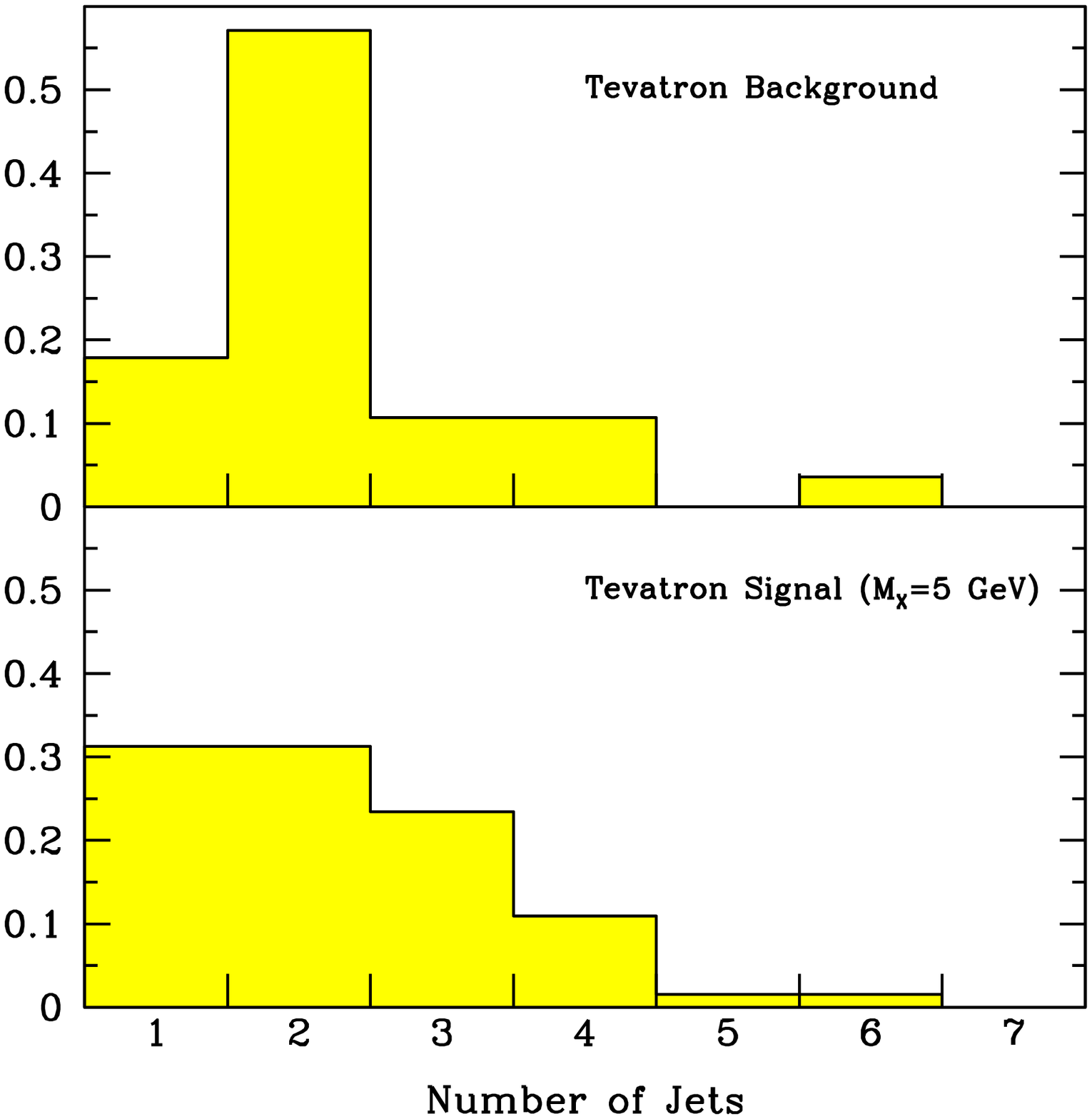}
\includegraphics[width=3.5in,angle=0]{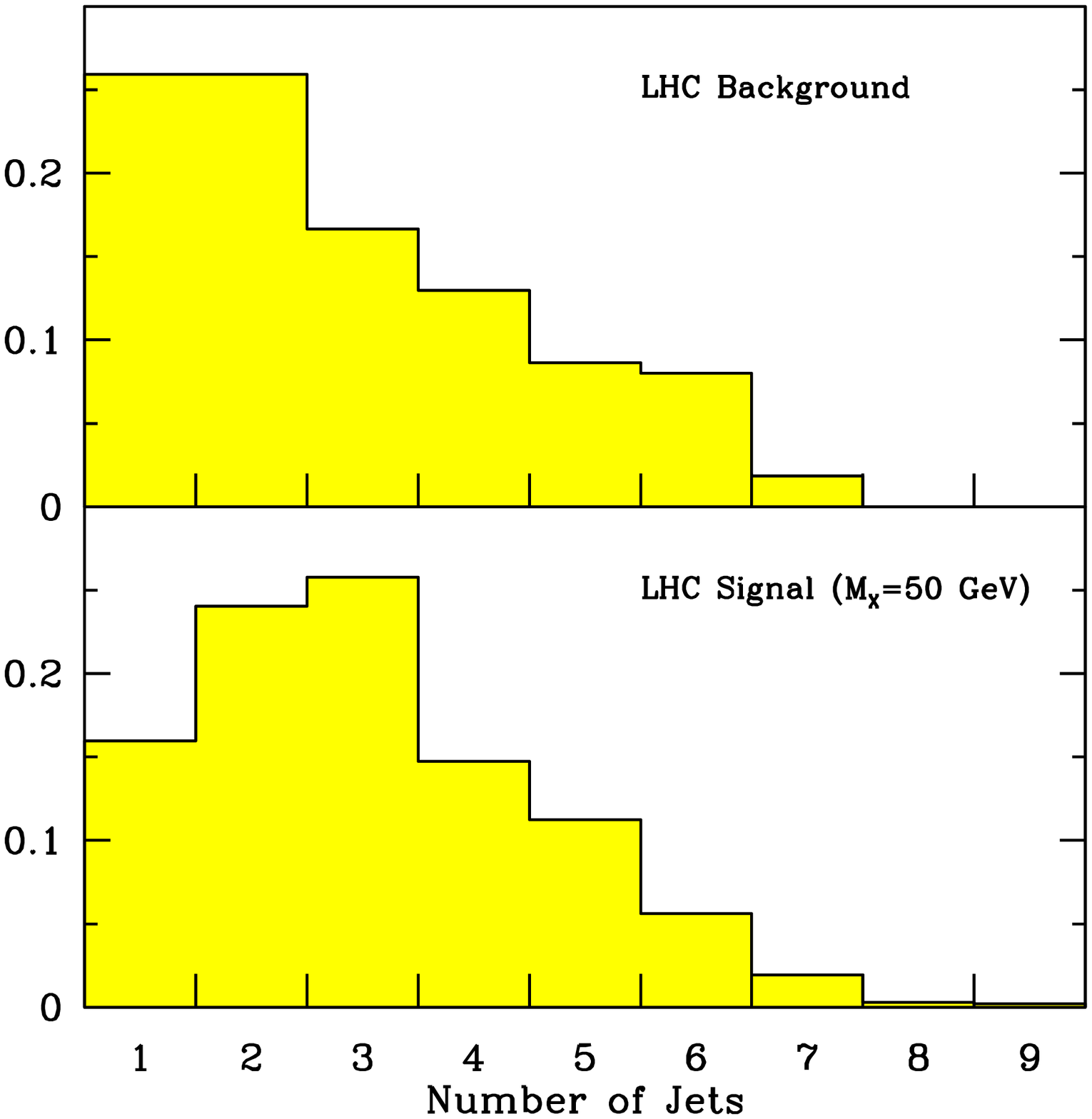}

\caption{Distributions for the number of jets (after the $p_T$ and
$\slashed{E}_T$ cuts discussed in the text) for signal and background processes
at the Tevatron for a WIMP mass of $5$ GeV  (left) and the LHC for a WIMP mass
of $50$ GeV (right).   For each case, the sum of the background processes are
shown in the upper panels, and the signal distributions in the lower panels.}

\label{jets}
\end{figure}

The precise manifestation of the harder partons in the signal events depends to
some degree on the details of the analysis.  Harder partons have more energy to
radiate, and produce a spray of more energetic particles, some of which may form
separate additional jets.  We find that for the cone algorithm (with cone size
$R = 0.4$), the signal events typically end up with three or more jets after
parton showering, and the the backgrounds peak at two (but both signal and
background have substantial tails).  The distribution of the number of jets of
both signal and background processes is shown in Fig.~\ref{jets}.  The higher
energies of the primary partons in the signal process reflects itself in a
higher jet multiplicity.  This in turn reflects itself in larger typical values
of the $H_T = \sum | \vec{p}_T | + \slashed{E}_T$ variable.

\begin{figure}
\centering
\includegraphics[width=7in,angle=0]{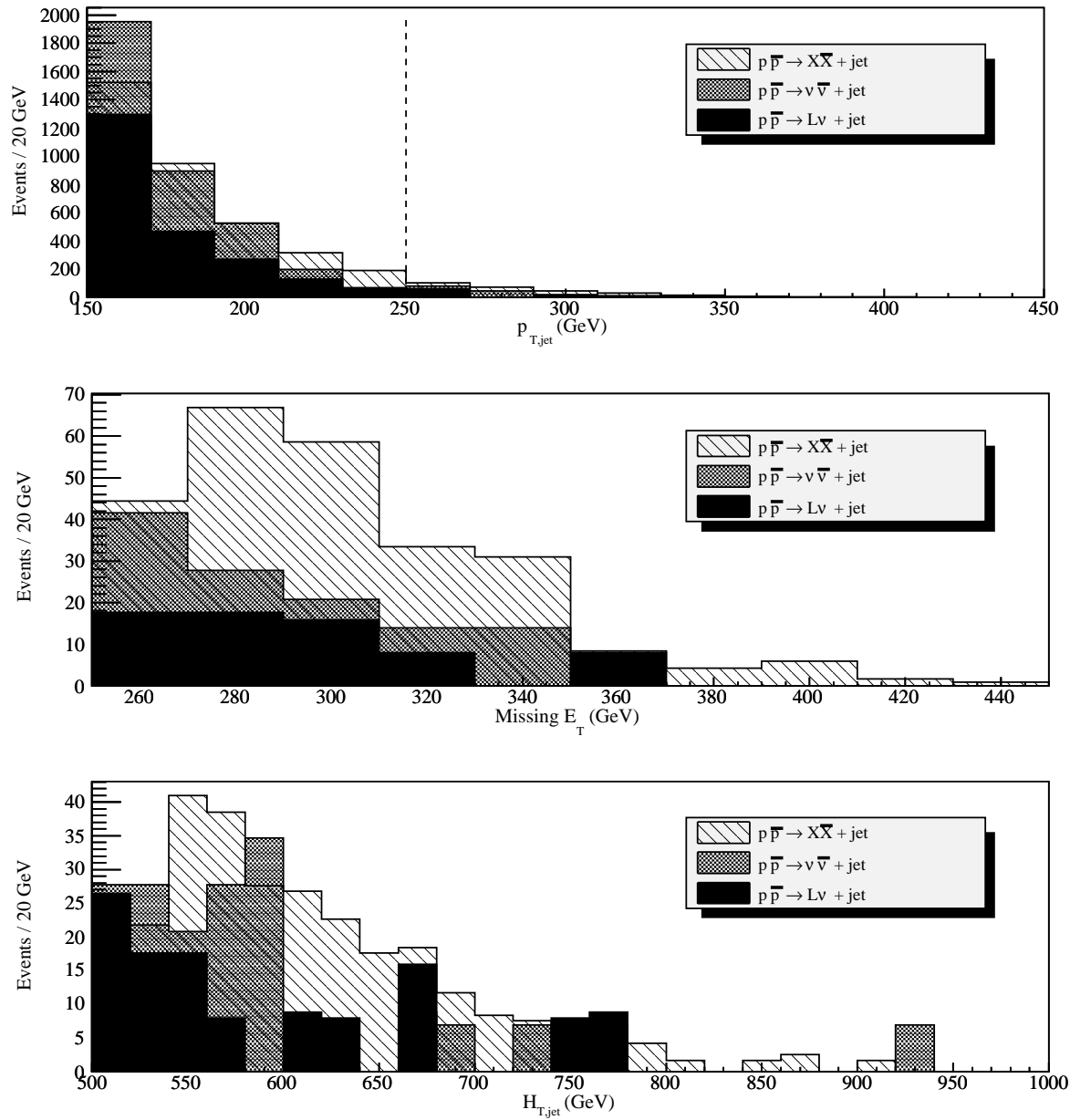}

\caption{The number of events as a function of $p_{T,\textrm{jet}}$ (top), $H_T$
(middle), and $\slashed{E}_T$ (bottom) for WIMP production and standard-model
background processes at the  Tevatron with $\sqrt{s} = 1.96$ TeV with $\int L\
\textrm{dt} = 10\ \textrm{fb}^{-1}$, and a WIMP mass $M_X=5$ GeV. The dotted
line in the upper panel indicates the cut we impose on $p_{T,\textrm{jet}}$.}

\label{tev196pthtmet}
\end{figure}

\begin{figure}
\centering
\includegraphics[width=7in,angle=0]{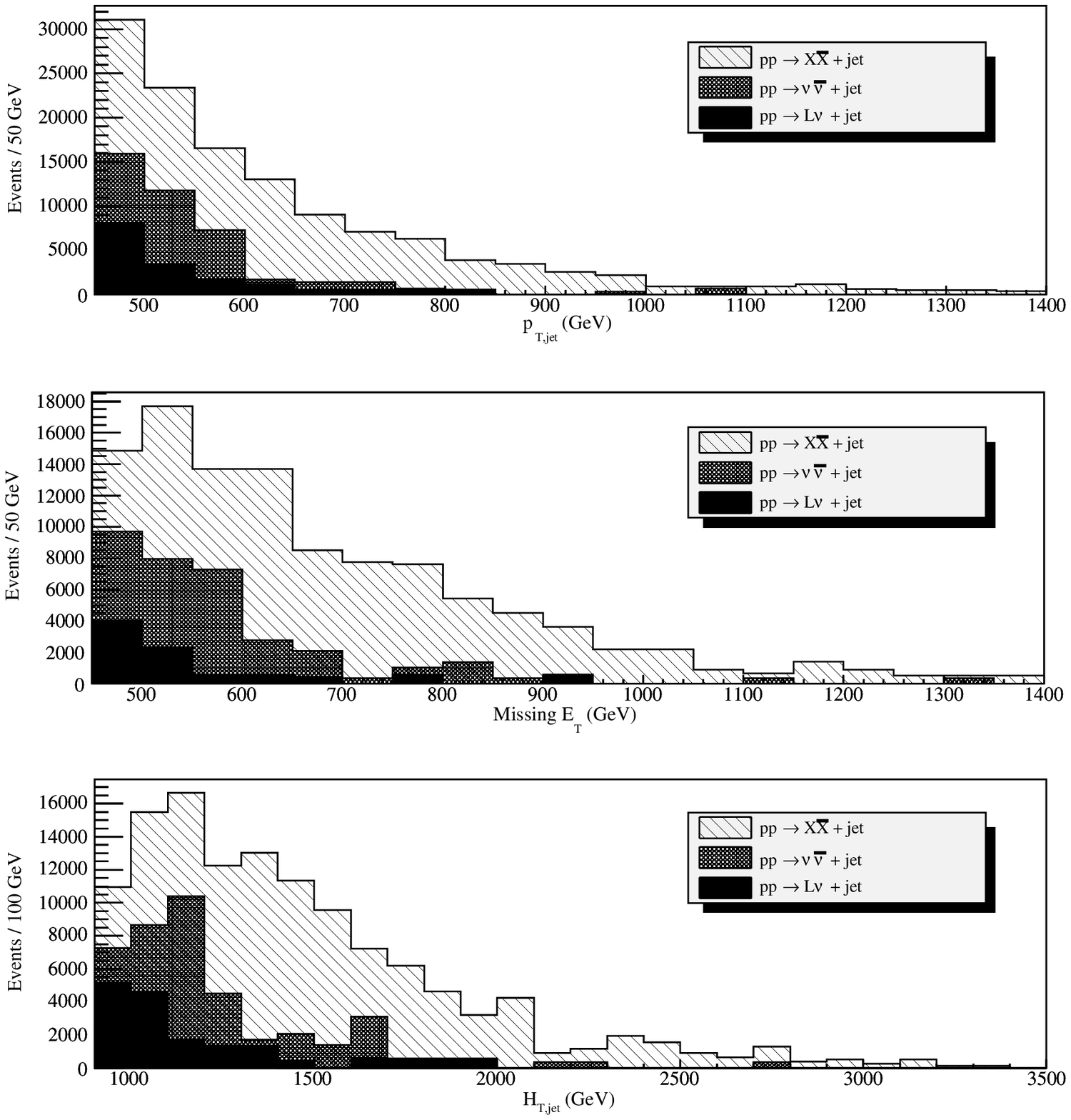}

\caption{The number of events as a function of $p_{T,\textrm{jet}}$ (top), $H_T$
(middle), and $\slashed{E}_T$ (bottom) for WIMP production and the most
important standard-model background processes at the LHC  with $\sqrt{s} = 14$
TeV with $\int L\ \textrm{dt} = 100\ \textrm{fb}^{-1}$, and a WIMP mass
$M_X=50$ GeV.}

\label{lhc14pthtmet}
\end{figure}

In Fig.\ \ref{tev196pthtmet} we present the distributions of the $p_T$ of the
leading jet, $H_T$, and $\slashed{E}_T$ for signal and background at the
Tevatron.  For the signal, we have chosen a WIMP mass of 5 GeV.  The coupling
$G_A$ is fixed such that $X$ reproduces the correct thermal relic density, and
the normalization corresponds to an integrated luminosity of $\int L ~
\textrm{dt} = 10\ \textrm{fb}^{-1}$.  The position of our primary cut $p_T \geq
250$~GeV is indicated by the dashed line, and clearly provides very effective
separation of signal from background.  Fig.\ \ref{lhc14pthtmet} shows the same
three distributions for the $\sqrt{s} = 14$ TeV LHC, where the signal
corresponds to a WIMP of mass 50 GeV, and the integrated luminosity is $ \int L
~ \textrm{dt} = 100\ \textrm{fb}^{-1}$. 

\section{Projected Sensitivity}

Ideally, one would compare the different shapes of the various observables:
$p_T$ of the leading jet, $\slashed{E}_T$, and $H_T$ and/or the number of jets
between the collider data and the predictions for the SM backgrounds, and either
put a limit on a combination of the mass of the maverick dark-matter particle
and its interaction strength, or conclude that there is evidence for a signal of
physics beyond the Standard Model.  Such a shape-based analysis is beyond the
scope of this work.

To estimate the approximate reach, we treat the signal as a counting experiment,
and ask for which $M_X$ (with $G_A$ continuing to be set for each $M_X$ in order
to hold the thermal relic density at its measured value) the signal would be
large enough to represent a statistically significant deviation above the
background.  In practice, this is probably a reasonable approximation to the
shape analysis to which we alluded above.  Since we have seen that for high
enough leading jet $p_T$ the background falls rapidly while the signal falls
much more slowly, we expect that a rate analysis is much like a  crude shape
analysis with a single high $p_T$ bin.

The backgrounds may be controlled by using data themselves.  The $Z+$ jets
process can be studied through the visible $Z \rightarrow \ell^+ \ell^-$ decay
mode, which is relatively easy to reconstruct thanks to the charged leptons, and
allows one to measure the associated jet distributions and the $p_T$ of the $Z$
itself (which in the missing energy analysis is the $\slashed{E}_T$). 
Similarly, one can study the $W+$ jets process for cases in which the charged
lepton can be reconstructed, and extrapolate into the regions where the charged
lepton is missed.

From the signal and background rates (after cuts) in Table \ref{crosssections1},
we determine the expected number of signal events ($S$) and the sum of the
background events ($B$) expected with integrated luminosities of 10
$\textrm{fb}^{-1}$ (at the Tevatron) and 100 $\textrm{fb}^{-1}$ (at the 14 TeV
LHC), as a function of the WIMP mass.  After cuts, the expected background rates
are still expected to produce 100 events or more, and Gaussian statistics may be
safely applied.  Requiring a signal which is significant at the $5\sigma$, $S /
\sqrt{B} \geq 5$.  We find that a five sigma deviation is expected for WIMP
masses,
\begin{align}
\textrm{Tevatron}:~& M_X \lesssim 15~\textrm{GeV} \nonumber \\
\textrm{14 TeV LHC}:~ & M_X \lesssim 275~\textrm{GeV}  . 
\end{align}
If no deviation is observed, $95\%$ C.L. limits may be obtained,
\begin{align}
\textrm{Tevatron}:~& M_X \gtrsim 25~\textrm{GeV} \nonumber \\
\textrm{14 TeV LHC}:~ & M_X \gtrsim 450~\textrm{GeV} . \footnotemark 
\end{align}
\footnotetext{The corresponding WIMP mass limits for an LHC run with $\sqrt{s} = 7$ TeV and a total
integrated luminosity of $\int L \textrm{dt} = 1$ fb$^{-1}$ are $M_X = 40$ GeV for a five-sigma detection,
and $M_X = 55$ GeV for a $95\%$ c.l.\ limit.}

One can also relax the assumption that $X$ is a thermal relic and determine
limits in the $M_X$--$G_A$ parameter plane.

\section{Conclusions}

In this paper we have explored the idea that the dark matter is a cold thermal
relic, and it is the only beyond the standard-model particle relevant for dark
matter that is within the reach of the LHC, \textit{i.e.,} a maverick particle.
We have employed a particular well motivated effective field theory description
of the WIMP interactions with standard-model particles---Eq.\
(\ref{interactionlagrangian})---which specifies the cosmological/astrophysical
annihilation, scattering with nuclei, and  collider production of the WIMP. 
While the formalism is general,  we explore the case of a Dirac WIMP interacting
through an axial-vector interaction, which leads to spin-dependent low-energy
WIMP-nucleon scattering, and consistent with the up until now null results of
direct-detection experiments \cite{beltran09}.  However, it is worth bearing in
mind that for WIMPs with masses greater than about 200 GeV, vector and scalar
interactions are also compatible with direct detection limits.   Since the
corresponding values of $G$ motivated by the relic density for   those cases are
smaller, the signals for a thermal relic will be   somewhat more challenging to
pick out from the backgrounds. It would be interesting to pursue those cases as
well in future work.

We have further chosen flavor-independent coupling to quarks, as might be
expected in a theory which does not lead to unacceptably large corrections to SM
flavor observables. In fact, provided the WIMP has significant coupling to the
up and down quarks which are present as valence quarks in the proton, we do not
expect relaxing that assumption to change our results very significantly. We
have also assumed no direct coupling to leptons.  If there were coupling to
leptons, then depending on the UV completion, there may also be new
contributions to processes such as $q \bar{q} \rightarrow \ell^+ \ell^-$.  Such
a process has excellent discovery prospects at the LHC, and would provide
information about the greater context in which the WIMP lives, but is
model-dependent and thus not part of a minimal signal of maverick dark matter.

Here, we explore the possibility of production and detection at colliders of a
pair of maverick WIMPs together with hard jets, $p p \ (p \bar{p}) \rightarrow X
\bar{X} + \textrm{jets}$.  We find that the associated jets tend to have larger
transverse momentum and missing transverse energy compared to background events,
providing potential handles that can be exploited to extract the signal from the
background. In particular, the numbers of jets, as well as differences in the
number of events as a function of leading jet transverse momenta,
$p_{T,\textrm{jet}}$, the scalar sums of jet transverse momenta plus missing
transverse energy, $H_T = \sum | \vec{p}_T | + \slashed{E}_T$, and the missing
transverse energy, $\slashed{E}_T$, differs between signal and background.  This
leads to a promising discovery potential for this class of maverick WIMPs. 
Because our WIMPs are produced through higher dimensional operators without the
appearance of intermediate on-shell colored particles, we are lead to devise
somewhat different analysis strategies than those typically employed for {\em
e.g.,} searches for supersymmetry or large extra dimensions.

We estimate the discovery mass reach of the Tevatron with 10 fb$^{-1}$ total
integrated luminosity to be about 15 GeV and the mass reach of the  14 TeV LHC
with  an integrated luminosity of 100 fb$^{-1}$  (roughly a year's running at
design luminosity) to be 275 GeV. These numbers are rough estimates, for there
are three issues we did not properly account for: 1) We have not seriously
attempted to optimize the cuts, which would require a more detailed detector
simulation, and more sophisticated simulation of the backgrounds (ideally using
data as input); 2) Using $S/\sqrt{B}$ as an indication of the mass reach is
reasonable for a counting experiment, but ideally one would perform a shape
analysis, making use of the fact that the background and signal lead to
different $p_T$ and $H_T$ distributions; and 3) We have ignored uncertainties in
the background calculation.

Properly taking into account issues 1) and 2) above would result in a better
limit, while issue 3) would weaken the limit.  To get some idea of the
uncertainty related to issue 3), we can compare to the CDF analysis in Ref.\
\cite{Abulencia:2006kk} based on 368 pb$^{-1}$ of $p\bar{p}$ collisions.  With
the cuts employed in their analysis, they expect a SM background of 265 $\pm$ 30
events (including 15 $\pm$ 10 QCD multijet events), and observe 263 events. 
Using the Bayesian approach of Ref.\ \cite{Heinrich:2004tj} results in a maximum
number of 67 signal events from new physics (including a 13\% uncertainty in the
signal acceptance).  Using the same cuts as in the CDF analysis, we would expect
67 maverick dark matter events for a mass of 14 GeV.  Our na\"{\i}ve
$S/\sqrt{B}$ analysis would result in a mass limit of 11 GeV.  This leads us to
believe that our projected limits may not be too far off what is achievable (and
in fact may be quite conservative).  It also suggests that the QCD multijet
background may not be an issue.

Ultimately, we expect that it will be difficult to push the mass reach beyond
15 GeV and 400 GeV, for the Tevatron and LHC respectively, because there is
simply not enough energy available to create pairs of more massive WIMPs.

Missing transverse energy searches have been proposed as a strategy to explore
several scenarios for BSM physics.  A question for our analysis (in fact, a
question for all WIMP searches at colliders) is that if a missing transverse
energy signal that cannot be described by standard-model physics is detected,
how do we know it is related to dark matter?  One possible answer comes in the
form of a consistency check: the production cross section  depends upon both the
mass of the WIMP, $M_X$, and its coupling to quarks, $G_A$.  If the maverick
WIMP is the dark matter, we can infer $G_A(M_X)$ from the relic density
calculation.  The kinematics of the event distribution (such as the peak in the
$p_{T,\textrm{jet}}$ distribution) carries information about $M_X$. One can ask
whether the combination of $G_A(M_X)$ for dark matter and $M_X$ from kinematical
determinations agrees with the measured total production cross
section.\footnote{If the two match, it will be highly suggestive of a discovery
of dark matter.  If they do not, one will be left wondering if the new signal of
missing energy is unrelated to dark matter, or if the WIMP is perhaps not a
thermal relic.}  In addition, if signals of direct or indirect detection become
evident, one can also correlate with them.  

A maverick scenario of dark matter is interesting, and we have demonstrated
that collider experiments can constrain (or discover!) maverick dark matter.


\acknowledgments{We would like to thank Florencia Canelli, John Conway, 
Jonathan Feng, Henry Frisch, JoAnne Hewett, Ben Kiliminster, 
Tom LeCompte, Frank Petriello,
Tilman Plehn, Will Shepherd, Jay Wacker, and Lian-tao Wang  for  useful
discussions.  This  work was supported in part by the Department of Energy at
the University of Chicago and Fermilab.  T.\ Tait is grateful to the SLAC theory
group for their hospitality during his many visits.}




\end{document}